\documentclass[12pt]{iopart}
\usepackage{graphicx}
\usepackage{enumerate}
\graphicspath{{Figurer/}}

\renewcommand{\em}{\mathbf{e}} 				% \em allerede defineret

\newcommand{\Am}{\mathbf{A}}
\newcommand{\am}{\mathbf{a}}

\newcommand{\Fm}{\mathbf{F}}
\newcommand{\Nm}{\mathbf{N}}
\newcommand{\fm}{\mathbf{f}}
\newcommand{\Rm}{\mathbf{R}}
\newcommand{\rma}{\mathbf{r}} 				% renew \rm virker ikke
\newcommand{\vm}{\mathbf{v}}
\newcommand{\G}{\mathbf{G}}

\newcommand{\M}{\mathcal{M}}
\renewcommand{\L}{\mathcal{L}}
\newcommand{\I}{\mathcal{I}}
\newcommand{\Q}{\mathcal{Q}}
\newcommand{\D}{\Delta}

\newcommand{\ti}{\theta_i}
\newcommand{\ta}{\theta}

\begin{document}

\title[Swinging counterweight trebuchet]
{The swinging counterweight trebuchet\\
On internal forces}

\author{E Horsdal}
\address{Department of Physics and Astronomy, Aarhus University,
DK-8000 Aarhus C, Denmark}
\ead{horsdal@phys.au.dk}

\begin{abstract}
	The forces that act internally in a trebuchet as it delivers a shot	depend on the
	motions of throwing arm, counterweight and sling.
	These motions are considered known experimentally or theoretically
	and given	in the form of time-dependent angular coordinates.
	Explicit expressions in terms of these coordinates and their derivatives
	to second order are derived for the internal forces.
	The forces that act immediately after a shot is initiated can be extracted from 
	the equations of motion without solving them, and they are compared with 
	static forces just prior to initiation.	
	Required strengths of the different parts of a trebuchet depend on the 
	internal forces, which also determine sliding friction losses. 
	Illustrative results are given for a specific trebuchet.
\end{abstract}
\section{Introduction}
The coupled differential equations that govern the internal movement of a trebuchet with 
swinging counterweight must be derived from mechanical energies without reference to 
unknown internal forces. 
These remain unknown even after the equations have been integrated, 
but once the movement is established, theoretically or experimentally, the internal forces 
can be calculated by the use of Newton's second law.

The internal forces determine the strengths of the various parts of the trebuchet
and the inevitable loss of mechanical energy to sliding friction:
The pivoting shaft of the throwing arm must be able to carry the heavy load from
the counterweight when its fall is suddenly interrupted, 
and so must the hinge by which it is attached to the arm.
Heat in proportion to load is generated at the bearings, and this loss
of mechanical energy reduces range and kinetic energy at target in addition to 
inflicting wear and degradation of the wood.
Also, the projectile is accelerated in a sling with two cords, and the tension of each
rises to values much larger than just half the gravity of the projectile.
The cords must be sufficiently strong to withstand the tension, and this 
also applies to the spigot and ring that holds the sling.
The bending load from sling tension and counterweight strains the throwing arm 
and may even break it.
The engine's supporting structure has the shape of a trestle with high-positioned bearings.
The trestle must be sufficiently strong and heavy to prevent it from deforming, sliding, 
or tilting due to the forces on the bearings, 
which have large components both horizontally and vertically.
\section{Trebuchet and angular coordinates}
\begin{figure}[htb]
\centering
	\includegraphics[width=0.85\textwidth]{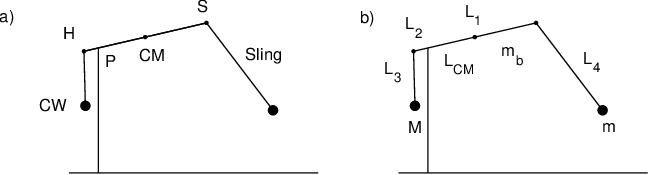}
	\caption{
		Trebuchet.
		a) Moving parts and fixed pivot.
		b) Lengths and masses.}
	\label{fig:Config}
\end{figure}

Schematic diagrams of a trebuchet are shown in~\fref{fig:Config}a and~\ref{fig:Config}b.
It consists of three moving parts identified in~\ref{fig:Config}a: 
A throwing arm HS supported at a fixed pivoting point P, a counterweight CW free to swing 
about a hinge H, and a sling for the projectile attached at S.
The throwing arm, also referred to as the beam, is treated as a rigid body
with mass~$m_b$, a fixed pivoting point~$P$, 
a center of mass located a distance~$L_{CM}$ from~$P$, 
and a moment of inertia~$\I$ relative to~$P$.
It is divided by~P into long and short segments of lengths~$L_1$ and~$L_2$, 
respectively, as seen in~\fref{fig:Config}b.
The counterweight of mass~$M$ is treated as a point particle 
placed at the end of a weightless arm of length~$L_3$, which 
is hinged to the throwing arm at~H.
The sling of length~$L_4$ is attached to the arm at the spigot~S, 
and the projectile of mass~$m$ is also treated as a point particle.

A shot runs through three phases:
The projectile is in the sling and drawn along the bottom of a trough at ground 
level during phase I.
It is lifted off the trough at the start of phase~II and remains in the sling until 
it is released into a ballistic trajectory.
This marks the beginning of phase~III 
that lasts until the engine comes to rest.

\begin{figure}[htb]
	\centering
	\includegraphics[width=0.85\textwidth]{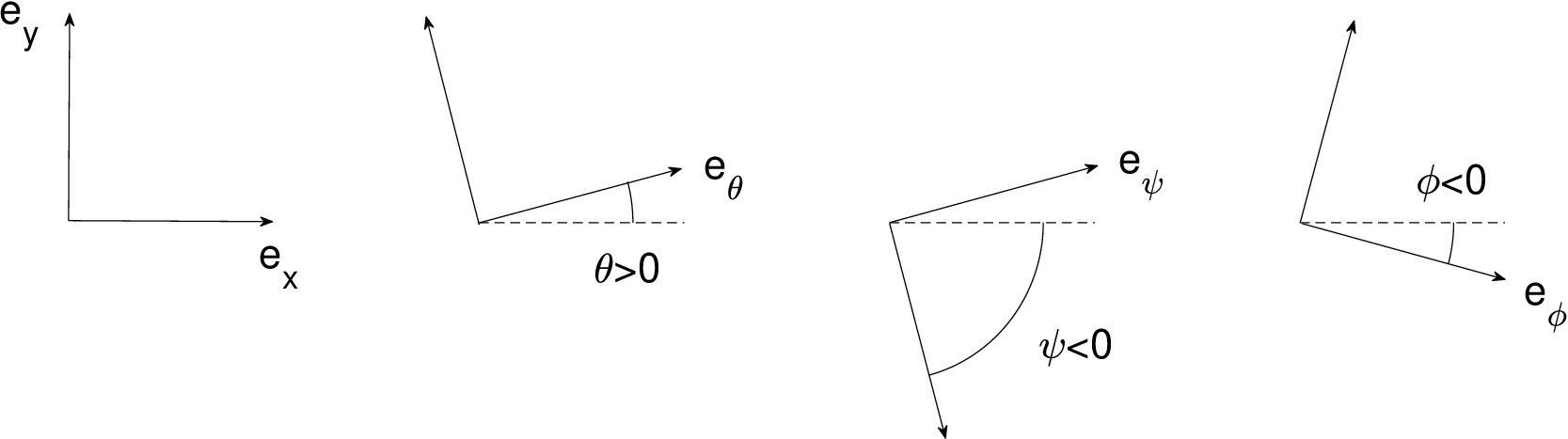}
	\caption{Angular coordinates and unit vectors in the directions of beam~$\em_\ta$, 
		counterweight~$\em_\psi$ and projectile~$\em_\phi$.
		Fixed unit vectors~$(\em_x,\em_y)$.}
	\label{fig:UnitVec}
\end{figure}

The kinematics of beam, counterweight and projectile is described by 
the angles~$\theta$,~$\psi$ and~$\phi$, respectively,
which are shown in~\fref{fig:UnitVec}. 
A fixed coordinate system and three that follow the motions are also shown.
The unit vectors~$\em_x$ and~$\em_y$ define the fixed system, and with reference 
to~\fref{fig:Config}, the unit vector~$\em_\theta$ points from P to S,~$\em_\psi$ 
from H to M, and~$\em_\phi$ from S to m.

\Tref{tab:Dimensions} gives an example of linear dimensions, masses and initial
angles for a large engine inspired by drawings in the sketchbook of Villard de 
Honnecourt~\cite{ref:VdH}.
It is optimized to throw~100kg stones a distance of~450m in vacuum 
(the range in air is~435m~\cite{ref:VirTreb}).
The moment of inertia is~$\I=m_b(L_1^2-L_1L_2+L_2^2)/3$, and
the optimization procedure, discussed in~\cite{ref:Horsdal}, 
is designed such that it limits internal forces, minimizes the mass 
of the throwing arm, and ensures a large efficiency above~90\% with
an available mechanical energy~$\D U=255$kJ.
\begin{table}[tbh]\footnotesize
	\centering
		\begin{tabular}{ccc|cc|ccc|c|ccc}
			\multicolumn{5}{c|}{Lengths} 							&
			\multicolumn{3}{c|}{Masses} 							&	
			\multicolumn{1}{c|}{MI} 									&	
			\multicolumn{3}{c}{Initial angles} 				\\\hline
			\multicolumn{3}{c|}{Beam} 								&	
			\multicolumn{1}{c}{CW} 										&
			\multicolumn{1}{c|}{Sling} 								&
			\multicolumn{1}{c}{Beam} 									&
			\multicolumn{1}{c}{CW} 										&
			\multicolumn{1}{c|}{Stone} 								&
			\multicolumn{1}{c|}{Beam} 								&
			\multicolumn{1}{c}{Beam} 									&
			\multicolumn{1}{c}{CW} 										&
			\multicolumn{1}{c}{Sling} 								\\
			Long & Short & CM & arm & & &  & & & &  &	\\
			 $L_1$ & $L_2$ & $L_{CM}$ & $L_3$ & $L_4$ & $m_b$ 		& $M$ 	& $m$ & $\I$ & $\ti$ & $\psi_i$&  $\phi_i$	\\\hline
			 m & m & m & m & m & kg 		& kg 	&kg & kgm$^2$ & & &\\
			 6.48	& 0.86	& 2.81 & 2.45	& 5.55	&622	& 19100	& 100 & 7704 & $-50^\circ$&$-90^\circ$&$-180^\circ$
		\end{tabular}
	\caption{A large trebuchet. Lengths, masses, moment of inertia (MI) and initial angles.
	Initial angular speeds and accelerations equal zero.
	CW, counterweight. 
	CM, center of mass for beam.}
	\label{tab:Dimensions}
\end{table}
\section{Kinematics}\label{sec:EoM_Kin}
The unit vectors~$(\em_\theta,\em_\psi,\em_\phi)$ in~\fref{fig:UnitVec} 
follow the motion of 
beam, counterweight and projectile, respectively.
In cartesian coordinates they are
\begin{eqnarray}\nonumber%\label{eq:UnitVec_1}
		\em_a
		=\cos(a)\em_x+\sin(a)\em_y	
		,
\end{eqnarray}
where~$a$ is~$\ta$,~$\psi$ or~$\phi$.
The perpendicular vectors  of the coordinate systems are
\begin{eqnarray}\nonumber%\label{eq:UnitVec_2}
		\em_{a\perp}
		=-\sin(a)\em_x+\cos(a)\em_y 
		~.
\end{eqnarray}
They depend on time and the derivatives are
\begin{eqnarray}
	\dot\em_a&=\dot a\em_{a\perp},
	\quad\quad
	&\ddot\em_a=\ddot a\em_{a\perp}-\dot a^2\em_a,
	\nonumber\\
	\dot\em_{a\perp}&=-\dot a\em_a,
	\quad\quad
	&\ddot\em_{a\perp}=-\ddot a\em_a+\dot a^2\em_{a\perp},\nonumber
\end{eqnarray}
where we use Newton's dot notation for differentiation.

\begin{itemize}
\item Projectile in phase I:\\	
	The position is
	\begin{eqnarray}\label{eq:rP}
		\rma_m=H\em_y+L_1\em_\ta+L_4\em_\phi,
	\end{eqnarray}
	where~$H=-L_1\sin\ti$ is the height of P in~\fref{fig:Config}, and	
	the projectile slides in the trough, 
	so there is a bond between~$\ta$ and~$\phi$ given by~$\rma_m\cdot\em_y=0$, or
	\begin{eqnarray}\label{eq:Const_I1}
		H+L_1\sin\ta+L_4\sin\phi=0.
	\end{eqnarray}
	The angular speeds are also related
	\begin{eqnarray}\label{eq:Const_I2}
		L_1\cos\ta~\dot\ta+L_4\cos\phi~\dot\phi&=&0.					
	\end{eqnarray}
	Differentiation of~\eref{eq:rP} and use of~\eref{eq:Const_I1}
	and~\eref{eq:Const_I2} leads to the velocity~$\vm_m$ and acceleration~$\am_m$
	\begin{eqnarray}
	 \vm_m&=&
	 L_1f(\ta)\dot\ta\em_x
	 \label{eq:vP_I}							\nonumber\\
	 \am_m&=&
	 L_1
	 \left(
	 f(\ta)\ddot\ta+g(\ta)\dot\ta^2
	 \right)\em_x									
	 ,\label{eq:aP_I}
	\end{eqnarray}
	where 
	\begin{eqnarray}
		f(\ta)&=-\frac{\sin(\ta-\phi)}{\cos\phi} \qquad\mathrm{and}			\label{eq:f_def}\\
		g(\ta)&=
		-\frac{L_1}{L_4}\frac{\cos^2\ta}{\cos^3\phi}
		-\frac{\cos(\ta-\phi)}{\cos\phi}																\label{eq:g_def}
		~.
	\end{eqnarray}
\item Projectile in phase II:\\
	Position is given by~\eref{eq:rP}, 
	and velocity and acceleration are, respectively,
	\begin{eqnarray}
	 \vm_m&=&
	 L_1\dot\ta\em_{\ta\perp}+L_4\dot\phi\em_{\phi\perp}
	 \label{eq:vP}																							
	\hspace{40pt}\mathrm{and}															\nonumber\\
	 \am_m&=&
	 L_1(\ddot\ta\em_{\ta\perp}-\dot\ta^2\em_\ta)+
	 L_4(\ddot\phi\em_{\phi\perp}-\dot\phi^2\em_\phi).		\nonumber
	\label{eq:aP}
	\end{eqnarray}
\item Counterweight in all phases:\\
	Position
	\begin{eqnarray}\nonumber%\label{eq:rC}
		\rma_M=H\em_y-L_2\em_\ta+L_3\em_\psi,
	\end{eqnarray}
	and velocity~$\vm_M$ and acceleration~$\am_M$
	\begin{eqnarray}
	 \vm_M&=&
	 -L_2\dot\ta\em_{\ta\perp}+L_3\dot\psi\em_{\psi\perp}
	 \nonumber\\
	 \am_M&=&
	 -L_2(\ddot\ta\em_{\ta\perp}-\dot\ta^2\em_\ta)+
	 L_3(\ddot\psi\em_{\psi\perp}-\dot\psi^2\em_\psi).
	\label{eq:aCW}
	\end{eqnarray}
\item Center of mass for beam in all phases:\\
	Position
	\begin{eqnarray}\nonumber
		\rma_{CM}=H\em_y+L_{CM}\em_\ta,
	\end{eqnarray}
	and velocity~$\vm_{CM}$ and acceleration~$\am_{CM}$
	\begin{eqnarray}
		\vm_{CM}&=&
		L_{CM}\dot\ta\em_{\ta\perp}								\nonumber\\
		\am_{CM}&=&
		L_{CM}(\ddot\ta\em_{\ta\perp}-\dot\ta^2\em_\ta)
		\label{eq:aCM}.
	\end{eqnarray}
\end{itemize}	
\section{Mechanical energies}
\begin{itemize}
\item Projectile:\\	
	Kinetic and potential energies are, respectively,
	\begin{eqnarray}\nonumber
		T_m=\frac{1}{2}m\vm_m^2
		\hspace{44pt}\mathrm{and}\hspace{30pt}
		U_m=mg\rma_m\cdot\em_y.
	\end{eqnarray}
	In phase~I
	\begin{eqnarray}
		T_m&=
		\frac{1}{2}m\left(L_1f(\ta)\dot\ta\right)^2								\nonumber\\		
		U_m\quad&=0.\nonumber
	\end{eqnarray}
	In phase~II
	\begin{eqnarray}
		T_m&=
		\frac{1}{2}m(L_1\dot\ta\em_{\ta\perp}+L_4\dot\phi\em_{\phi\perp})^2										\nonumber\\
		&=
		\frac{1}{2}m(L_1^2\dot\ta^2+L_4^2\dot\phi^2+2L_1L_4\dot\ta\dot\phi\cos(\ta-\phi))		\nonumber\\
		U_m\quad&= 
		mg(L_1\em_\ta+L_4\em_\phi)\cdot\em_y  \nonumber\\
		&=mg(L_1\sin\ta+L_4\sin\phi).  \nonumber
	\end{eqnarray}
\item Counterweight in both phases:
	%%
	%\begin{eqnarray}\nonumber
		%T_M=\frac{1}{2}M\vm_M^2
		%\hspace{30pt}\mathrm{and}\hspace{30pt}
		%U_M=Mg\rma_M\cdot\em_y.
	%\end{eqnarray}
	%%
	%In both phases
	%
	\begin{eqnarray}
		T_M&=
		\frac{1}{2}M(L_2^2\dot\ta^2+L_3^2\dot\psi^2+2L_2L_3\dot\ta\dot\psi\cos(\ta-\psi))		\nonumber\\
		U_M\quad&=Mg(-L_2\sin\ta+L_3\sin\psi).  \nonumber
	\end{eqnarray}
\item Beam in both phases:
	\begin{eqnarray}\nonumber
		T_{m_b}=\frac{1}{2}\I\dot\ta^2
		\hspace{48pt}\mathrm{and}\hspace{30pt}
		U_{m_b}=m_bgL_{CM}\sin\ta
	\end{eqnarray}
	where~$\I$ is the moment of inertia for rotation around the pivoting axle.
\item 
	Lagrange function and equations of motion.\\	
	The total kinetic and potential energies are
	\begin{eqnarray}\nonumber
		T=T_m+T_M+T_{m_b}
		\qquad\mathrm{and}\qquad
		U=U_m+U_M+U_{m_b}.
	\end{eqnarray}
	The equations for the internal movement of the trebuchet are derived from the 
	Lagrange function~$\L=T-U$ by the use of the Lagrange equations.
\item 
	Energy invested in loading.\\	
	This is the difference~$\D U$ of potential energies in initial and final configurations.
	The angular coordinates in the initial configuration are~$\ta=\ta_i$ and~$\psi_i=-\pi/2$, 
	and in the final~$\ta_f=\pi/2$ and~$\psi_f=-\pi/2$, so
	\begin{eqnarray}\nonumber
		\D U=(ML_2-m_bL_{CM})g(1-\sin\ta_i).
	\end{eqnarray}
\end{itemize}
\section{Static initial forces}\label{sec:SF}
We first look at the conditions when the engine is ready to be fired and
all parts are at rest.
The forces on the beam at the hinge~H and at the center of mass~CM are then~$-Mg\em_y$ 
and~$-m_bg\em_y$, respectively.
The static reaction~$\Fm_R$ from the bearings at~P, on which the pivoting shaft 
of the beam rests, depends on how the beam is prevented from rotating.
We consider two possibilities.
\begin{enumerate}[a)]
\item
	A locking force~$\Fm_L=-F_L\em_{\ta\perp}$ is applied perpendicular to the beam 
	at the spigot.
	The magnitude of the force~$F_L$ is such that the torque relative to the pivot
	vanishes and therefore
	\begin{eqnarray}\nonumber%\label{eq:Frp}
		F_L=\frac{(ML_2-m_bL_{CM})g\cos\ti}{L_1}
		.
	\end{eqnarray}
	The total force on the beam also vanishes, so the reaction force~$\Fm_a$ satisfies
	\begin{eqnarray}\nonumber
		\Fm_a-Mg\em_y-m_bg\em_y+\Fm_L&=0,
	\end{eqnarray}
	and this leads to 
	\begin{eqnarray}
		\Fm_a&=(\Fm_a\cdot\em_x)\em_x + (\Fm_a\cdot\em_y)\em_y             \nonumber \\
		&=-F_L\sin\ti\em_x+
		\Big(
		(M+m_b)g+
		F_L\cos\ti
		\Big)\em_y.\nonumber
	\end{eqnarray}
	The projectile just lies in the trough and the tension of the sling is zero.
\item
	Another possibility is to hold the projectile with a firm grip such that the locking force
	is horizontal and applied through the sling.
	The magnitude of this locking force~$F_L$ now satisfies
	\begin{eqnarray}\nonumber%\label{eq:Frh}
		F_L=\frac{(ML_2-m_bL_{CM})g\cos\ti}{L_1\sin\ti}
		,
	\end{eqnarray}
	and the reaction force is
	\begin{eqnarray}\nonumber
		\Fm_b=
		-F_L\em_x+(M+m_b)g\em_y
		.
	\end{eqnarray}
\end{enumerate}
\section{Dynamic forces}\label{sec:EoM_F}
\begin{figure}[tbh]
\centering
	\includegraphics[width=0.50\textwidth]{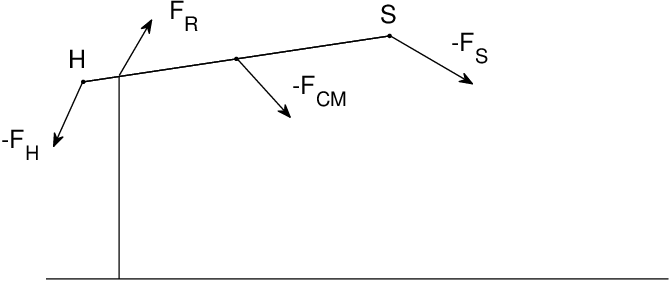}
	\caption{Forces on the beam~$-\Fm_H$,~$-\Fm_{CM}$,~$-\Fm_S$ and~$\Fm_R$.}
	\label{fig:Forces}
\end{figure}

Physical forces originating from the rigid beam act on the counterweight and projectile,
and they are~$\Fm_H$ for the counterweight and~$\Fm_S$ for the projectile.
The force on the center of mass is~$\Fm_{CM}$.
The negative of these forces act on the arm and they are shown 
in~\fref{fig:Forces}
at the points where they attack.
There is also a reaction force~$\Fm_R$ on the beam at the pivoting point~P.
This is at rest, so the sum of the four forces equals zero,
\begin{eqnarray}\label{eq:F_R}
	\Fm_R-\Fm_H-\Fm_{CM}-\Fm_S=\bf{0}.
 \label{eq:FR}
\end{eqnarray}

The total dynamic forces on the counterweight, center of mass and projectile 
are~$\Fm_H-Mg\em_y$,~$\Fm_{CM}-m_bg\em_y$ and~$\Fm_S-mg\em_y$, respectively,
and they can all be calculated by Newtons' second law when the motion is known.
Subtraction of gravity then determines~$\Fm_H$,~$\Fm_{CM}$ and~$\Fm_S$ and
finally the reaction force~$\Fm_R$ by~\eref{eq:F_R}.
  
To simplify calculations and expressions, it is convenient to use matrix notation 
for the unit vectors in~\fref{fig:UnitVec}
\begin{eqnarray}\nonumber
\em_x=
	 \left\{			
	 \begin{array}{c}
			1 \\
			0		 
	 \end{array}
	 \right\},~
\em_y=
	 \left\{			
	 \begin{array}{c}
			0 \\
			1		 
	 \end{array}
	 \right\},~
\em_a=
	 \left\{			
	 \begin{array}{c}
			\cos(a) \\
			\sin(a)		 
	 \end{array}
	 \right\},~
\em_{a\perp}=
	 \left\{			
	 \begin{array}{c}
			-\sin(a) \\
			\cos(a)		 
	 \end{array}
	 \right\},
\end{eqnarray}
and to introduce a rotation matrix~$\Rm_a$	
and a generalized angular acceleration~$\Am_a$
\begin{eqnarray}\label{eq:RA}
	 \Rm_a=
	 \left\{			
	 \begin{array}{cc}
			\cos(a) & -\sin(a) \\
			\sin(a) &  \cos(a)		 
	 \end{array}
	 \right\}, 
	\qquad
	\Am_a=
	 \left\{			
	 \begin{array}{c}
			-\dot a^2  \\
			\ddot a		 
	 \end{array}
	 \right\}.
\end{eqnarray}	
The dynamic forces can then be expressed as sums of terms, 
which each has the appearance of mass times acceleration:~$M_aL_a\Rm_a\Am_a$.
The acceleration~$L_a\Rm_a\Am_a$ is rotated from~$L_a\Am_a$, 
but the two accelerations have equal magnitudes.

The forces vary most dramatically and are strongest during phase~II. 
We treat this first, then return to phase~I, and continue in~\sref{sec:Torque} 
with expressions for the torque on the throwing arm and in~\sref{sec:IAF} 
with initial discontinuities of forces.
\subsection{Phase II}\label{sec:EoM_FII}
\begin{itemize}
\item Center of mass for beam:\\	
	The total force on the imaginary center of mass particle of the throwing arm is~$m_b\am_{CM}$.
	This is the sum of gravity~$-m_bg\em_y$ and a physical force~$\Fm_{CM}$ from within 
	the rigid arm, so
	\begin{eqnarray}\nonumber%\label{eq:FCM}
		m_b\am_{CM}=\Fm_{CM}-m_bg\em_y
	\end{eqnarray}
	The acceleration~$\am_{CM}$ was given in~\eref{eq:aCM}.
	$\Fm_{CM}$ therefore takes the form
	\begin{eqnarray}\label{eq:FCMxy} 							
		\Fm_{CM}&=m_bL_{CM}\left(-\dot\ta^2\em_\ta+\ddot\ta\em_{\ta\perp}\right)+m_bg\em_y  \nonumber\\
		&=
		m_bL_{CM}\Rm_\ta\Am_{\ta}+m_bg\em_y,
	\end{eqnarray}
	where~$\Rm$ and~$\Am$ are defined in~\eref{eq:RA}.		
	The internal force~$\Fm_{CM}$ has a component along the beam that tends to 
	stretch or compress it, and a perpendicular component that tends to bend it.
	Both components contribute to the reaction force at the pivot. 
\item Counterweight and hinge:\\	
	The total force on the counterweight is~$M\am_M$.
	This is the sum of gravity~$-Mg\em_y$ and the physical force from the hinge~$\Fm_H$,
	\begin{eqnarray}\nonumber%\label{eq:FCW}
		M\am_M=\Fm_H-Mg\em_y.
	\end{eqnarray}
	The acceleration~$\am_M$ was given in~\eref{eq:aCW}, so~$\Fm_H$ takes the form
	\begin{eqnarray}\label{eq:FHxy}
		\Fm_H=-ML_2\Rm_\ta\Am_{\ta}+ML_3\Rm_\psi\Am_{\psi}+Mg\em_y,
	\end{eqnarray}
	The vector~$\Fm_H$ is parallel to~$\em_\psi$.
	This is necessarily the case in experiments because the arm for 
	the counterweight could	be a flexible string, 
	and the equations of motion also ensure~$\Fm_H\cdot\em_{\psi\perp}=0$.
\item Projectile and spigot:\\
	The total force on the projectile is~$m\am_m$.
	This is the sum of gravity~$-mg\em_y$ and the physical sling tension~$\Fm_S$, so
	\begin{eqnarray}\label{eq:SlingForce}
		m\am_m=\Fm_S-mg\em_y
	\end{eqnarray}
	and therefore
	\begin{eqnarray}\label{eq:FSxy}
		\Fm_S=mL_1\Rm_\ta\Am_{\ta}+mL_4\Rm_\phi\Am_{\phi}+mg\em_y.
	\end{eqnarray}
	with~$\am_m$ from~\eref{eq:aP}.	
	The vector~$\Fm_S$ is parallel to~$\em_\phi$.
\item Reaction on pivoting axle:\\
	The reaction force~$\Fm_R$ is given by~\eref{eq:F_R},~\eref{eq:FCMxy},~\eref{eq:FHxy} 
	and~\eref{eq:FSxy}, and one finds
\begin{eqnarray}\label{eq:FR_xy}
	 \Fm_R=&
	 \M_0g\em_y+\nonumber\\
	 &\M_1\Rm_\ta\Am_\ta+
	 ML_3\Rm_\psi\Am_\psi+
	 mL_4\Rm_\phi\Am_\phi,
\end{eqnarray}	
where all term that depend on~$\ta$ are collected in one, and
\begin{eqnarray}\nonumber%\label{eq:M0M1}
	\M_0=m+M+m_b
	\quad\mathrm{and}\quad
	\M_1=mL_1-ML_2+m_bL_{CM}
\end{eqnarray}
are~1st and~2nd moments of the masses, respectively.
The first term in~\eref{eq:FR_xy} is the constant gravity
and the three remaining depend on each angular motion.
\begin{enumerate}
\item Components of~$\Fm_R$ along beam and perpendicular:\\
		The components of~$\Fm_R$ in the rotating basis~$(\em_\ta,\em_{\ta\perp})$ where
		the throwing arm is at rest	are found by applying the rotation~$\Rm_{-\ta}$, so
		\begin{eqnarray}%\label{eq:FR_beam}
			\Fm_{R\ta}&=\Rm_{-\ta}\Fm_R \nonumber\\
			&=\M_0\Rm_{-\ta}\G+	\nonumber\\		
			&\quad~\M_1\Am_\ta+
			ML_3\Rm_{\psi-\ta}\Am_\psi+
			mL_4\Rm_{\phi-\ta}\Am_\phi,\nonumber
		\end{eqnarray}	
		where
		\begin{eqnarray}\nonumber
			 \G=g
			 \left\{			
			 \begin{array}{c}
			 		0 \\ 1		 
			 \end{array}
			 \right\}.
		\end{eqnarray}	
\item Strength of~$\Fm_R$:\\
		The magnitude~$F_R$ of~$\Fm_R$ can be found from~\eref{eq:FR_xy}	
		by multiplying~$\Fm_R$ by itself
		\begin{eqnarray}
			F_R^2&=(\Fm_R)^T\Fm_R \nonumber\\
			&=(\M_0g)^2+
			\left(\M_1\Am_\ta\right)^2+
			\left(ML_3\Am_\psi\right)^2+
			\left(mL_4\Am_\phi\right)^2			 	\nonumber\\
			&+
			2\M_0\G^T
			\left(
			\M_1\Rm_\ta\Am_\ta+
			ML_3\Rm_\psi\Am_\psi+
			mL_4\Rm_\phi\Am_\phi
			\right)															\nonumber\\
			&+
			2\M_1ML_3
			\Am_\psi^T\Rm_{\ta-\psi}\Am_\ta 		\nonumber\\
			&+
			2\M_1mL_4
			\Am_\phi^T\Rm_{\ta-\phi}\Am_\ta 		\nonumber\\
			&+
			2ML_3mL_4
			\Am_\phi^T\Rm_{\psi-\phi}\Am_\psi.	\nonumber
		\end{eqnarray}
		The first four terms are the direct ones,
		and the remaining six are twelve cross terms combined two by two.
		The terms are scalars so unchanged when transposed,
		which eliminates an apparent asymmetry.
	\end{enumerate}
\end{itemize}
\subsection{Phase I}\label{sec:EoM_FI}
\begin{itemize}
\item Projectile.\\
	The projectile slides in the trough with acceleration in the horizontal 
	direction only
	\begin{eqnarray}\nonumber%\label{eq:FS_p}
		\am_m=
		L_1
	 \left(
	 f(\ta)\ddot\ta+g(\ta)\dot\ta^2
	 \right)\em_x,
	\end{eqnarray}
	where~\eref{eq:aP_I} is used and the functions~$f$ and~$g$
	are defined in~\eref{eq:f_def} and~\eref{eq:g_def}.
	The force~$\Fm_S$ is necessarily parallel to the sling, so~$\Fm_S\cdot\em_{\phi\perp}=0$,
	and it drives the projectile, so~$(\Fm_S\cdot\em_x)\em_x=m\am_m$.
	These conditions on~$\Fm_S$ imply
	\begin{eqnarray}\label{eq:FS_I}
		\Fm_S=
		mL_1\left(f(\ta)\ddot\ta+g(\ta)\dot\ta^2\right)(\em_x+\tan\phi\em_y).
	\end{eqnarray}
\item Counterweight and center of mass of beam.\\	
	The forces~$\Fm_H$ and~$\Fm_{CM}$ are as in 
	phase II,~\textit{i.e.}~\eref{eq:FCMxy} and~\eref{eq:FHxy}, 
	respectively.
\item Reaction on pivoting axle.\\
	The reaction force is~$\Fm_R=\Fm_{CM}+\Fm_H+\Fm_S$, 
	and with~$\fm_\ta=\left\{-g(\ta),f(\ta)\right\}$ it reads
	\begin{eqnarray}\label{eq:FR_init}
		\Fm_R
		&=
		(M+m_b)g\em_y+mL_1\fm_\ta\Am_\ta(\em_x+\tan\phi\em_y)+	 	\nonumber\\
		&\quad~(m_bL_{CM}-ML_2)\Rm_\ta\Am_\ta+ML_3\Rm_\psi\Am_\psi.
	\end{eqnarray}
\end{itemize}
\section{Torque}\label{sec:Torque}
\begin{figure}[tbh]
\centering
	\includegraphics[width=0.60\textwidth]{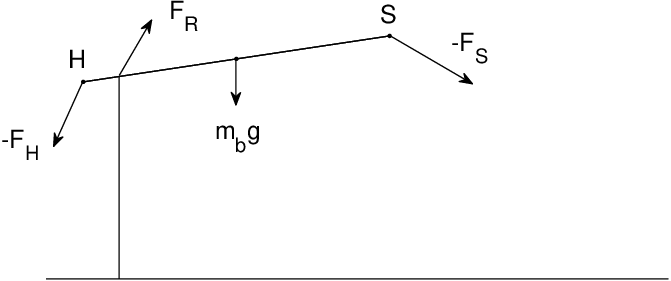}
	\caption{External forces on beam.}
	\label{fig:ForcesRot}
\end{figure}

The throwing arm is treated as a rigid body that rotates under the influence
of the external forces shown in~\fref{fig:ForcesRot}.
These are~$-\Fm_H$ at hinge,~$\Fm_R$ at fulcrum,~$-m_bg\em_y$ at center of mass, 
and~$-\Fm_S$ at spigot.
We consider motion relative to the fulcrum, so the contribution from~$\Fm_R$
vanishes and therefore (with the form~$\Nm=\sum\rma\times\Fm$) 
\begin{eqnarray}
	\Nm
	&=-L_2\em_\ta\times(-\Fm_H)+L_{CM}\em_\ta\times(-m_bg\em_y)+L_1\em_\ta\times(-\Fm_S),	\nonumber\\
	&=L_2(\em_\ta\times\Fm_H)-L_{CM}m_bg(\em_\ta\times\em_y)-L_1(\em_\ta\times\Fm_S), 					\nonumber\\
	&=L_2(\Fm_H\cdot\em_{\ta\perp})\em_z-L_{CM}m_bg(\em_y\cdot\em_{\ta\perp})\em_z-
					L_1(\Fm_S\cdot\em_{\ta\perp})\em_z,																						\nonumber
\end{eqnarray}
where~$\em_z=\em_{\ta}\times\em_{\ta\perp}$.
The terms are parallel, so the magnitude of~$\Nm$ is
\begin{eqnarray}\label{eq:N}
	N=L_2(\Fm_H\cdot\em_{\ta\perp})-L_{CM}m_bg(\em_y\cdot\em_{\ta\perp})-
					L_1(\Fm_S\cdot\em_{\ta\perp}).
\end{eqnarray}
\section{Initial accelerations, forces and torque}\label{sec:IAF}
All parts of the trebuchet are at rest prior to~$t=0$ when a shot is initiated, and
the dynamics is analyzed at a time immediately after.
The angles and angular velocities are continuous at~$t=0$, 
but the angular accelerations change discontinuously from zero to the finite 
values~$\ddot\ta_i$,~$\ddot\psi_i$ and~$\ddot\phi_i$.
These values can be extracted without solution from the equations of motion.
When the initial angular accelerations are determined, initial linear accelerations, 
forces and torques follow.
The results are illustrated by an example.
\subsection{Initial angular accelerations}
From the equations for the internal movement in phase~I, see~\ref{app:EOM},
one finds
\begin{eqnarray}
	\ddot\ti&=
	\frac{(ML_2-m_bL_{CM})g\cos\ti}
			 {mL_1^2\sin^2\ti+ML_2^2\cos^2\ti+\I}~>0 
	\quad\mathrm{because}\quad ML_2>m_bL_{CM} 				\nonumber\\
	\ddot\psi_i&=
	-\frac{L_2}{L_3}\sin\ti\ddot\ti\hspace{95pt}>0
	\quad\mathrm{because}\quad-\pi/2<\ti<0						\nonumber\\
	\ddot\phi_i&=
	\frac{L_1}{L_4}\cos\ti\ddot\ti\hspace{104pt}>0.		\label{eq:AngAcc}
\end{eqnarray}
Limiting values for heavy counterweights
\begin{eqnarray}\nonumber
	\ddot\ti=\frac{1}{\cos\ti}\frac{g}{L_2},\quad
	\ddot\psi_i=-\tan\ti\frac{g}{L_3}\quad\mathrm{and}\quad 										
	\ddot\phi_i=\frac{L_1}{L_4}\frac{g}{L_2}.						
\end{eqnarray}
\subsection{Initial linear accelerations}
\Eref{eq:aP_I} and~\eref{eq:aCW} give for the projectile
and counterweight, respectively,
\begin{eqnarray}
	 \am_m=
		 -L_1\sin\ti\ddot\ti\em_x
		\hspace{32pt}\mathrm{with}\qquad
		\lim_{M\rightarrow\infty}\am_m=-\tan\ti\frac{L_1}{L_2}g\em_x		
		 \label{eq:api}
\end{eqnarray}
and
\begin{eqnarray}
	 \am_M=
		 -L_2\cos\ti\ddot\ti\em_y
		\qquad~\mathrm{with}\qquad
		\lim_{M\rightarrow\infty}\am_M=-g\em_y.	
		\label{eq:cwi}
\end{eqnarray}
A heavy counterweight thus accelerates the projectile by much more 
than just gravity~$g$ as in a free fall.
With the parameters in~\tref{tab:Dimensions}, the acceleration 
is~$\simeq9$ times larger, and
the acceleration of the counterweight approaches that of a free fall.
\subsection{Initial dynamic forces}
\Eref{eq:api} shows pure horizontal acceleration for the projectile. 
Gravity is therefore balanced by a normal reaction force from the trough, 
so the total forces on the projectile before and just after initiation of 
a shot are
\begin{eqnarray}\nonumber
	\Fm_m=m
	\left\{
		\begin{array}{cll}
			0  													&\mathrm{at}& t=0^- \\
			-L_1\sin\ti\ddot\ti\em_x>0	&\mathrm{at}& t=0^+ .
		\end{array}		
	\right.
\end{eqnarray}
The counterweight is accelerated only vertically according to~\eref{eq:cwi}, 
so total forces are
\begin{eqnarray}\nonumber
	\Fm_M=M
	\left\{
		\begin{array}{cll}
			0 													&\mathrm{at}& t=0^-\\
			-L_2\cos\ti\ddot\ti\em_y 		&\mathrm{at}& t=0^+.
		\end{array}		
	\right.
\end{eqnarray}

The reaction force at the spigot depends on the locking for~$t<0$, and at~$t=0^+$ it
follows from~\eref{eq:FS_I} and equals~$\Fm_m$,
\begin{eqnarray}\label{eq:FSi}
	\Fm_S=-mL_1\sin\ti\ddot\ti\em_x 
	\quad\mathrm{at}\quad t=0^+.
\end{eqnarray}
The reaction force at the center of mass equals gravity before a shot, 
and at~$t=0^+$ it follows from~\eref{eq:FCMxy},
\begin{eqnarray}\label{eq:FCMi}
	\Fm_{CM}=m_b
	\left\{
		\begin{array}{ccc}
			g\em_y 																																&\mathrm{at} & t=0^-\\
			-L_{CM}\sin\ti\ddot\ti\em_x+\left(g+L_{CM}\cos\ti\ddot\ti\right)\em_y	&\mathrm{at} & t=0^+.
		\end{array}		
	\right.
\end{eqnarray}
The reaction force at the hinge is first gravity, and at~$t=0^+$ it follows from~\eref{eq:FHxy},
\begin{eqnarray}\label{eq:FHi}
	\Fm_H=M
	\left\{
		\begin{array}{cll}
			g\em_y 																	&\mathrm{at}& t=0^-\\
			\left(g-L_2\cos\ti\ddot\ti\right)\em_y	&\mathrm{at}& t=0^+,
		\end{array}		
	\right.
\end{eqnarray}
where the x-component has vanished because~$L_2\sin\ti\ddot\ti+L_3\ddot\psi_i=0$, see~\eref{eq:AngAcc}.

The initial static reaction force at the bearings for~$t<0$ is given 
in~\sref{sec:SF}, and from the sum of~\eref{eq:FSi},~\eref{eq:FCMi} and~\eref{eq:FHi} 
or alternatively from~\eref{eq:FR_init} we find
\begin{eqnarray}%\label{eq:F_Ry}
	\Fm_R=&-(mL_1+m_bL_{CM})\sin\ti\ddot\ti\em_x  								\nonumber\\
	&+\left[(M+m_b)g+(m_bL_{CM}-ML_2)\cos\ti\ddot\ti\right]\em_y
	\quad\mathrm{at}\quad t=0^+.	\nonumber
\end{eqnarray}
\subsection{Initial torque}
The initial torque on the throwing arm at~$t=0^+$ is
\begin{eqnarray}\nonumber
	N_i	=(L_2M-L_{CM}m_b)g\cos\ti-(mL_1^2\sin^2\ti+ML_2^2\cos^2\ti)\ddot\ti,
\end{eqnarray}
where~\eref{eq:N},~\eref{eq:FSi},~\eref{eq:FCMi} and~\eref{eq:FHi} were used,
but from~\eref{eq:AngAcc} follows
\begin{eqnarray}\nonumber
	(L_2M-L_{CM}m_b)g\cos\ti=(mL_1^2\sin^2\ti+ML_2^2\cos^2\ti+\I)\ddot\ti,
\end{eqnarray}
and this implies
\begin{eqnarray}\nonumber%\label{eq:Ni}
	N_i	=\I\ddot\ti,
\end{eqnarray}
which is the expected result.
\subsection{Example}\label{sec:ex}
Static and initial dynamic forces are shown in~\tref{tab:InitialForces}.
They are derived for the engine in~\tref{tab:Dimensions} and for locking perpendicular to the beam.
The initial angular acceleration of the throwing arm~$\ddot\ti$ equals~5.79rad/s$^2$ in this case.
\begin{table}[tbh]\footnotesize
	\centering
		\begin{tabular}{c||cc|cc||cc|cc|cc|cc}
			\multicolumn{1}{c||}{Forces} 						&
			\multicolumn{2}{c|}{$\Fm_m/mg$} 				&
			\multicolumn{2}{c||}{$\Fm_M/Mg$} 				&
			\multicolumn{2}{c|}{$\Fm_H/Mg$} 				&
			\multicolumn{2}{c|}{$\Fm_{CM}/Mg$} 			&	
			\multicolumn{2}{c|}{$\Fm_L/Mg$} 				&		
			\multicolumn{2}{c}{$\Fm_R/Mg$} 					\\
			\multicolumn{1}{c||}{at $t=$}	& h 	  & v & h	& v 		& h	& v  	 & h			& v     & h 			& v  			& h 		& v  		\\\hline
			\multicolumn{1}{c||}{$0^-$}  	& 0    	& 0	& 0	& 0	  	& 0	& 1	   & 0			& 0.033	& -0.058 	& -0.049	& 0.058	& 1.08	\\
			\multicolumn{1}{c||}{$0^+$} 	& 2.93	& 0	& 0	& -0.33	& 0	& 0.67 & 0.041	& 0.067	& 0 			& 0 			& 0.057	& 0.74	
		\end{tabular}
	\caption{Static and dynamic forces.
					Horizontal h and vertical v components.}
	\label{tab:InitialForces}
\end{table}
We see that the total force on the projectile~$\Fm_m=\Fm_S$ equals zero 
when the projectile just lies in the trough before a shot,
but it changes to~$\Fm_m/(mg)=2.93\em_x$ as soon as the shot is initiated.
The total force on the counterweight~$\Fm_M=\Fm_H-Mg\em_y$ is zero at rest before a shot,
but changes discontinuously to~$-0.33Mg\em_y$.
The reaction force at the hinge~$\Fm_H$ that keeps the counterweight
at rest before a shot drops from~$Mg\em_y$ to~$0.67Mg\em_y$, 
and the reaction force at the center of mass~$\Fm_{CM}$ is 
first~$m_bg\em_y=0.033Mg\em_y$, but changes direction and the magnitude goes up.
The locking force~$\Fm_L$ vanishes at release.
The reaction force~$\Fm_R$ shows finite horizontal and
vertical components before and after initiation.
The vertical component dominates and drops immediately when the shot is 
released.

The forces that govern the initial angular acceleration of the throwing arm 
are~$-\Fm_H$ at the hinge,~$-\Fm_S$ at the spigot, but only gravity~$-m_bg\em_y$ 
at the center of mass.
The torque~$N_i$ follows from these forces or from~$\I\ddot\ta_i$, and it has the 
value~45kNm.

\begin{figure}[tbh]
\centering
	\includegraphics[width=0.70\textwidth]{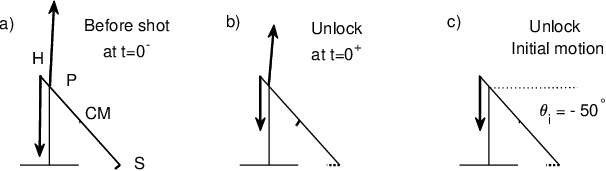}
	\caption{Forces at hinge H, pivot P, center of mass CM and spigot S.\\
						Small forces:
						$\Fm_{CM}=-m_bg\em_y$ in a) and c).
						$\Fm_S=-2.9mg\em_x$ in b) and c).}
	\label{fig:ForcesDisc}
\end{figure}

Directions and magnitudes of forces from~\tref{tab:InitialForces} are shown 
in~\fref{fig:ForcesDisc} as arrows.
The reaction forces in~\fref{fig:ForcesDisc}a and~\ref{fig:ForcesDisc}b
balance the forces at hinge, center of mass, and spigot.
The locking force in~\fref{fig:ForcesDisc}a applied perpendicular to the beam at the spigot 
changes to a very small horizontal force in~\fref{fig:ForcesDisc}b and~\ref{fig:ForcesDisc}c.
The vector that illustrates this force is rendered by a dotted line of length amplified by a factor
of~10 to make it visible. 
As seen in~\fref{fig:ForcesDisc}a and~\ref{fig:ForcesDisc}b, the apparent weight of the 
counterpoise and the load on the trestle decrease significantly at the start of a shot.
The external accelerating forces are shown in~\fref{fig:ForcesDisc}c, and they result in
an initial torque~$N_i$ of~45kNm. 
The contribution to~$N_i$ from the moving counterweight is~$0.67MgL_2\cos\ta_i=70$kNm,
so the small forces at the center of mass and spigot lower~$N_i$ by~36\%.
\section{Rates of work by forces} 
Work is done on the projectile, throwing arm and counterweight by the accelerating forces,
and this changes mechanical energies, 
but these changes can be evaluated also from the known motion, so the forces
are tested by a comparison.
\subsection{Projectile and counterweight}
The kinetic energy of the projectile is~$T_m=(1/2)m\vm_m^2$ and therefore
\begin{eqnarray}\nonumber%\label{eq:Fv}
	\frac{dT_m}{dt}=m\vm_m\cdot\frac{d\vm_m}{dt}=\Fm_m\cdot\vm_m.
\end{eqnarray}                                                       
There is also a gravitational potential 
energy~$U=-\int\Fm_gd\rma=mg\rma\cdot\em_y+U_0$,
which varies like
\begin{eqnarray}\nonumber%\label{eq:dUdt}
	\frac{dU_m}{dt}=mg\em_y\cdot\vm,
\end{eqnarray}
so the rate of change of the projectile's mechanical energy~$E_m=T_m+\D U_m$ is
\begin{eqnarray}\label{eq:dEdt}
	\frac{dE_m}{dt}=(\Fm_m+mg\em_y)\cdot\vm_m=\Fm_S\cdot\vm_m.
\end{eqnarray}
This is the rate of work done by the force~$\Fm_m$ plus the rate of work done 
against gravity~$\Fm_g$, or the rate of work done on the projectile by the sling tension~$\Fm_S$.
The mechanical energy accumulated by the projectile from time~$t=0$ to~$t$ is
\begin{eqnarray}\label{eq:Energy_m}
	E_m=\int_0^t\Fm_S\cdot\vm_mdt.
\end{eqnarray}
Likewise, power and accumulated mechanical energy for the counterweight are
\begin{eqnarray}\label{eq:Energy_M}
	\frac{dE_M}{dt}=\Fm_H\cdot\vm_M
	\qquad\mathrm{and}\qquad
	E_M=\int_0^t\Fm_H\cdot\vm_Mdt,
\end{eqnarray}
respectively.
\subsection{Throwing arm}
The throwing arm has the moment of inertia~$\I$ given in~\tref{tab:Dimensions} 
and the rotational kinetic energy is~$T_r=(1/2)\I\dot\ta^2$, so
\begin{eqnarray}\nonumber%
	\frac{dT_r}{dt}=\I\dot\ta\ddot\ta=N\dot\ta,
\end{eqnarray}
where~$N=\I\ddot\ta$ is the magnitude of the torque~$\Nm$ given in~\eref{eq:N}.
There is also a variation of potential energy, so for the rate of change of 
mechanical energy and accumulated energy we find, respectively
\begin{eqnarray}\label{eq:dUdt_beam}
	\frac{dE_a}{dt}=N\dot\ta+m_bg(\em_y\cdot\vm_{CM})
	\qquad\mathrm{and}\qquad
	E_a=\int_0^t\frac{dE_a}{dt}dt.
\end{eqnarray}
\section{Time dependencies}
For the engine specified in~\tref{tab:Dimensions},
the equations of motion were integrated numerically to determine the angular coordinates
as functions of time.
The next step, calculating forces, was not always taken by using expressions depending 
explicitly on these coordinates like equation~\eref{eq:FSxy} for the sling tension~$\Fm_S$.   
Instead, accelerations were most often found by numerical differentiation of positions, 
and forces then follow.
For~$\Fm_S$, the acceleration is~$d^2\rma_m/dt^2$ and the force is then given by~\eref{eq:SlingForce}.

The figures in this section show time-dependent internal forces and their loci, 
which are curves traced out by the end point of the vectors that represent the forces.
The torque on the throwing arm is also shown, as are the configurations of the trebuchet at the 
times when the bending force on the throwing arm is greatest, and when the projectile is released 
to achieve the best performance of the engine.
Figures that expose the instantaneous rates of work done on counterweight, 
throwing arm and projectile illustrate the transfer of mechanical energy within the engine.
\subsection{Forces}
\begin{figure}[htb]
\centering
	\includegraphics[width=0.55\textwidth]{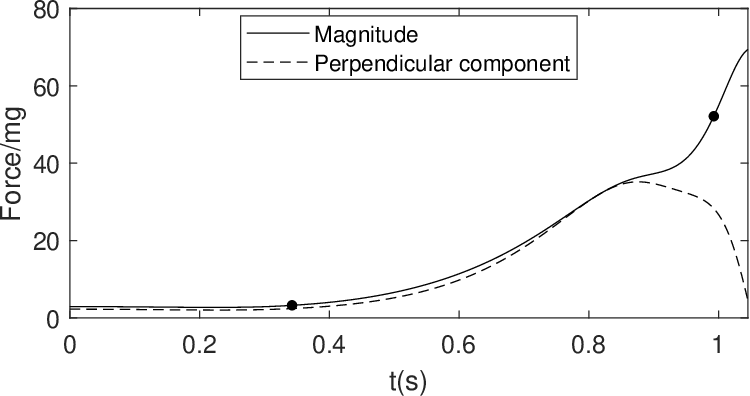}
	\caption{
		Tension of sling, half on each of two cords. 
		Magnitude and component perpendicular to beam.
		Full points at lift-off and release.}
	\label{fig:Tension}
\end{figure}

The sling tension~$\Fm_S$ has the magnitude~$F_S$ and the component perpendicular to the 
throwing arm is~$F_{S\perp}$. 
These quantities, measured in units of the projectile 
gravity~$mg$, are shown in~\fref{fig:Tension}.
The shot starts at~$t=0$, the projectile is lifted from the trough at~343ms, and the best time 
for release of the projectile is at~997ms, close to where the curves end.
The tension~$F_S$ at release jumps discontinuously to the initial dynamic value of~$2.93mg$ at~$t=0$, 
see~\tref{tab:InitialForces}, and stays almost constant during phase~I, but shortly into phase~II
it starts increasing strongly and reaches~$52mg$ at release.
With~$m=100$kg this amounts to the weight of~2.6 metric tonnes on each cord.
The component perpendicular to the beam is also shown.
It tends to bend the beam and shows a maximum value near~$35mg$ at~$876$ms.

\begin{figure}[htb]
\centering
	\includegraphics[width=0.90\textwidth]{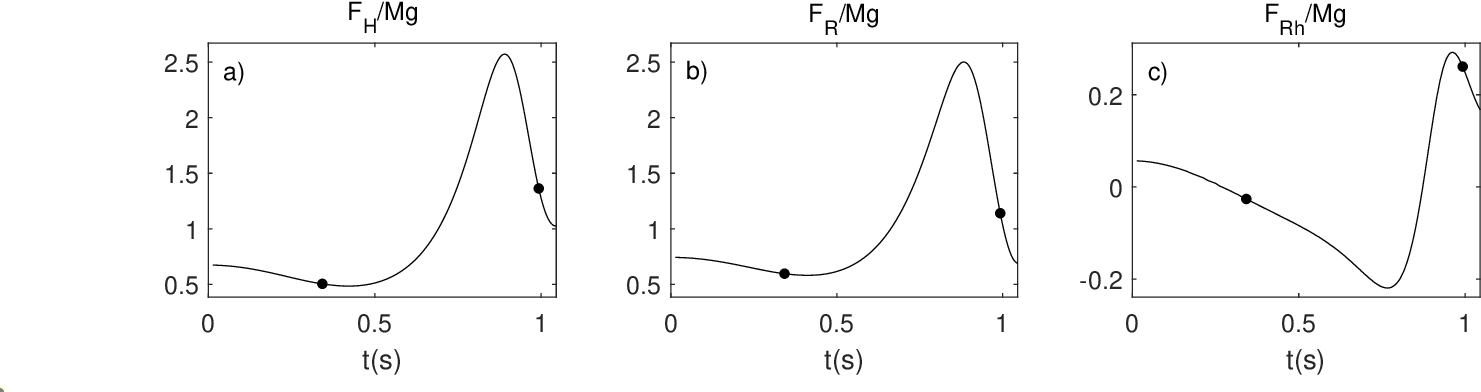}
	\caption{
		Reaction forces. 
		a) magnitude at hinge for counterweight, and
		b) at fulcrum of beam.
		c) horizontal component at fulcrum.
		Full points at lift-off and release.}
	\label{fig:Reaction}
\end{figure}

\Fref{fig:Reaction} shows reaction forces at hinge and fulcrum.
The force~$\Fm_H$ at the hinge in~\ref{fig:Reaction}a is first small in comparison with 
the gravity of the counterweight, and~\tref{tab:InitialForces} shows that~$F_H=0.67Mg$ 
at~$t=0^+$.
Thereafter, it first decreases to~$\simeq0.5Mg$ about halfway through the shot, then
rises to~$2.58Mg$, and is still larger than~$Mg$ at release.
The force at the fulcrum~$F_R$ shown in~\ref{fig:Reaction}b is not much different from this. 
It is a little larger at first but does not rise as high later on.
The horizontal component of~$F_R$, that tends to tilt and move the frame of the engine,
is first relatively small and pointing opposite the shooting direction, 
but soon passes through zero and then goes through an extremum of~$\simeq0.2Mg$ 
in the shooting direction before it again passes through zero and eventually reached 
a maximum of~$\simeq0.3Mg$ opposite the shooting direction just before release. 
This is what the trestle must be constructed to withstand.

\begin{figure}[htb]
\centering
	\includegraphics[width=0.95\textwidth]{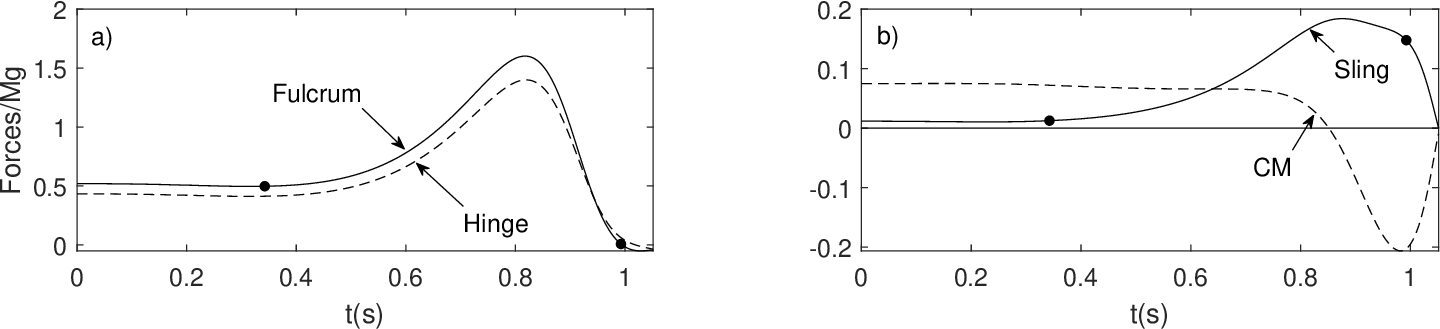}
	\caption{
		Reaction forces. Components perpendicular to beam.}
	\label{fig:PerpReac}
\end{figure}

Magnitudes of reaction forces perpendicular to the throwing arm are shown in~\Fref{fig:PerpReac}.
The arm is treated as a rigid body, but it is bend in practice 
and may even break at the pivot.
The bending load is the perpendicular reaction force at the 
fulcrum shown in~\fref{fig:PerpReac}a and 
given by~$\Fm_R\cdot\em_{\ta\perp}=(\Fm_H+\Fm_{CM}+\Fm_S)\cdot\em_{\ta\perp}$. 
The term~$\Fm_H$ dominates, and~\fref{fig:PerpReac}b shows that near the maximum 
of~$1.6Mg$, the sling contributes by~$\simeq10\%$ and the center of mass by much less.
\subsection{Loci of forces}
\begin{figure}[htb]
\centering
	\includegraphics[width=0.70\textwidth]{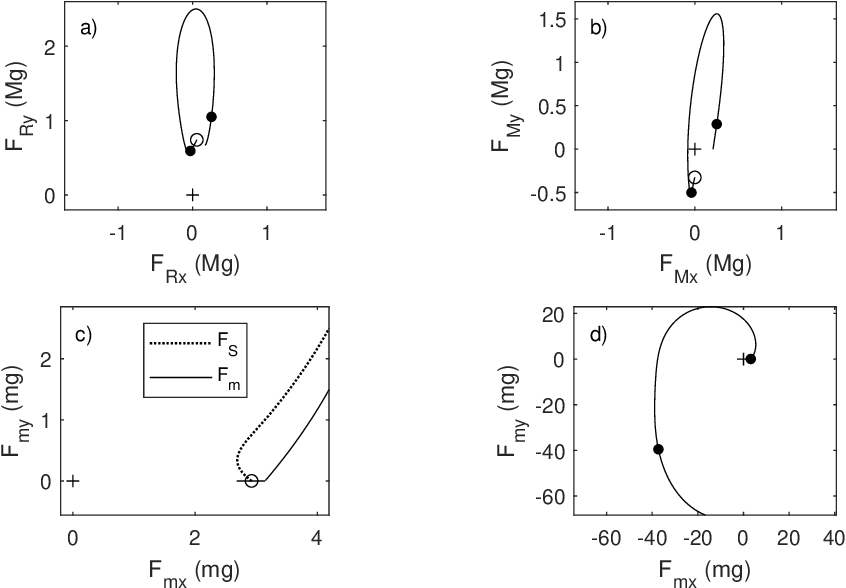}
	\caption{
		Forces during a shot.
		a)~$\Fm_R$.
		b)~$\Fm_M$.
		c)~$\Fm_m$ and~$\Fm_S$, initial.
		d)~$\Fm_m$.}
	\label{fig:ForceLocus}
\end{figure}

The reaction force~$\Fm_R$ that carries the throwing arm is seen in~\fref{fig:ForceLocus}a.
The open circle marks the start of the locus at~$t=0^+$ with magnitude less than~$Mg$, and it
drops further throughout phase~I from the open to the nearby closed circle. 
The magnitude thereafter increases strongly and the direction shifts to the right.
The magnitude is a little larger than~$Mg$ when the projectile is released at the second closed circle.

The total force on the counterweight~$\Fm_M$ is shown in~\fref{fig:ForceLocus}b.
The magnitude~$F_M$ is first zero, but jumps at~$t=0^+$ to~1/3 of the gravitational weight, 
and keeps increasing until lift-off where it reaches~$0.5Mg$. 
Hereafter, it decreases and the vector direction changes from 
mostly downwards to horizontal at~$t=695$ms.
The acceleration is now relatively small, but the motion is fast from the initial 
downwards acceleration.
Shortly after,~$F_M$ rises to a maximum larger than~$1.5Mg$ and now pointing mostly up.
The acceleration is then strong and against the motion.
This starts transforming the initial fall into an oscillatory motion characteristic of the 
behavior in phase~III. 
At release,~$F_M$ is close to~$0.5Mg$ with almost equal horizontal and vertical components.

The force on the projectile~$\Fm_m$ and the sling tension~$\Fm_S$ are related 
by the expression~$\Fm_m=\Fm_S-mg\em_y+F_N\em_y$, where the normal reaction~$F_N$ decreases 
during phase~I and vanishes at the end. 
The initial variations of~$\Fm_m$ and~$\Fm_S$ are shown in~\fref{fig:ForceLocus}c.
Both forces equal~$2.93mg\em_x$ at~$t=0^+$ (marked by an open circle) and the magnitude of 
the tension~$F_S$ thereafter first decreases and then increases while the vertical component 
keeps rising until it reaches the value~$mg$ and the projectile is lifted.
The projectile force is then~$\Fm_m=\Fm_S-mg\em_y$, and its increase and varying
direction during phase~II is shown in~\fref{fig:ForceLocus}d.
The largest lifting force goes beyond~$23mg$ and the force at release is more than twice as big.
\subsection{Torque}
\begin{figure}[htb]
\centering
	\includegraphics[width=0.60\textwidth]{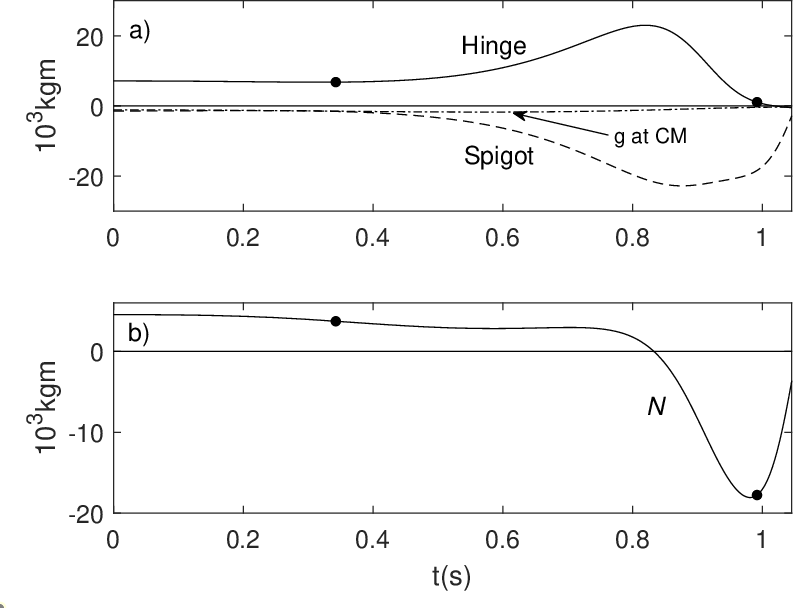}
	\caption{
		a) Contributions to torque from hinge, gravity at center of mass, and spigot.
		b) Full torque~$N$ on throwing arm.
		Full points at lift-off and release.}
	\label{fig:Torque}
\end{figure}

The torque~N is given in~\eref{eq:N} as the sum of three terms.
The contribution from each is shown in~\fref{fig:Torque}a.
The term that relates to the hinge is positive at all times up to release, 
and the one from the spigot is always negative.
The term from gravity at the center of mass is also negative throughout, but
contributes very little.
The total torque~$N$ is shown in~\fref{fig:Torque}b. 
It could have been calculated by~$N=\I\ddot\ta$, but this would hide the contribution 
from each term.
The torque is dominated by~$\Fm_H$ at first, but later on,~$\Fm_S$ contributes more 
and after~$\simeq850$ms, it dominates and slows down the rotation of the beam effectively.
The perpendicular reaction force at the pivot shown in~\fref{fig:PerpReac} goes through 
a maximum near~820ms and~$N$ goes through zero at almost the same time.
The maximum bending of the arm is thus due to almost equal moments of force at hinge and spigot.
$N$ is near a negative extremum at the time of release.
\subsection{Configurations}
\begin{figure}[htb]
\centering
	\includegraphics[width=0.65\textwidth]{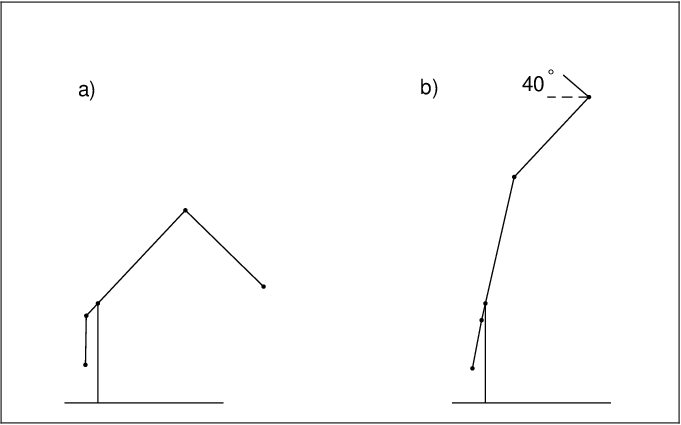}
	\caption{
		a) Configuration at maximum bending of throwing arm.
		b) Configuration at release.}
	\label{fig:Configurations}
\end{figure}

One can imagine that the throwing arm is supported horizontally at the ends, and loaded
at the pivoting point by a vertical force of magnitude~$|\Fm_R\cdot\em_{\ta\perp}|$.
The curvature and strain of the arm is then largest at the pivot, where it may
break if the strain exceeds a certain limit.
The maximum load read from~\fref{fig:PerpReac} is~$\simeq1.6Mg$ or~31 metric tonnes.  
The configuration of the trebuchet at this critical time is shown 
in~\fref{fig:Configurations}a.

The configuration of the trebuchet at~$t=997$ms, when release of the projectile leads to
maximum quality factor~$\Q$ for the given range and projectile mass~\cite{ref:Horsdal}, 
is shown in~\fref{fig:Configurations}b.
The throwing arm and the arm for the counterweight are almost parallel 
at this instant, so the torque from the counterweight has nearly vanished, 
and the counterweight is near its lowest position. 
The initial climb of the ballistic projectile motion is~$40^\circ$.
It starts~15.1m over ground and~5.55m behind the fulcrum at a speed of~66m/s or~238km/h.
The kinetic and potential energies are then~218kJ and~15kJ, respectively, and
they add to a kinetic energy of~233kJ at a horizontal target 
when internal friction and aerodynamic drag are ignored.
Experiments indicate that friction may reduce the mechanical energy by~$\simeq5\%$ 
and range by half as much~\cite{ref:EFJ}.
The effect of aerodynamic drag was calculated on the assumption of a spherical 
stone projectile with diameter~$D=0.42$m using the VirtualTrebuchet~2.0 
calculator~\cite{ref:VirTreb}, which leads to a reduction of range by~$\simeq3.3\%$ 
and of energy by twice that.
When added, the estimated reduction of range is by~$\simeq6\%$ and of energy 
by~$\simeq12\%$.

%of~450m according to 
%the VirtualTrebuchet~2.0 calculator~\cite{ref:VirTreb}.
%The relative loss of mechanical energy is about twice as big so the kinetic energy
%at target is reduced to 217kJ.
%
\subsection{Rate of work on projectile}
\begin{figure}[htb]
\centering
	\includegraphics[width=0.55\textwidth]{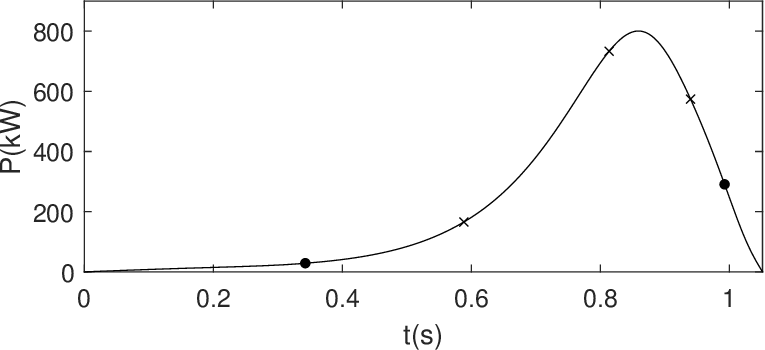}
	\caption{
		Rate of work done on projectile.
		Full points at lift-off and release.\\
		Crosses at~10\%,~50\% and~90\% of integrated power at release.}
	\label{fig:Power}
\end{figure}

The rate of work on the projectile by the sling force~$\Fm_S$ was calculated by the
use of~\eref{eq:dEdt} and is shown in~\fref{fig:Power}. 
The power is seen to be relatively small until well into phase~II,
but it eventually goes through a maximum larger than~800kW at~$t=859$ms.
The accumulated mechanical energy at the time of release is~233kJ.
More energy could be transferred to the projectile by releasing it later, and
this would increase the engine efficiency, but decrease range 
and the engine quality factor~$\Q$ defined in~\cite{ref:Horsdal}.
\subsection{Details on rates of work}
\begin{figure}[htb]
\centering
	\includegraphics[width=0.95\textwidth]{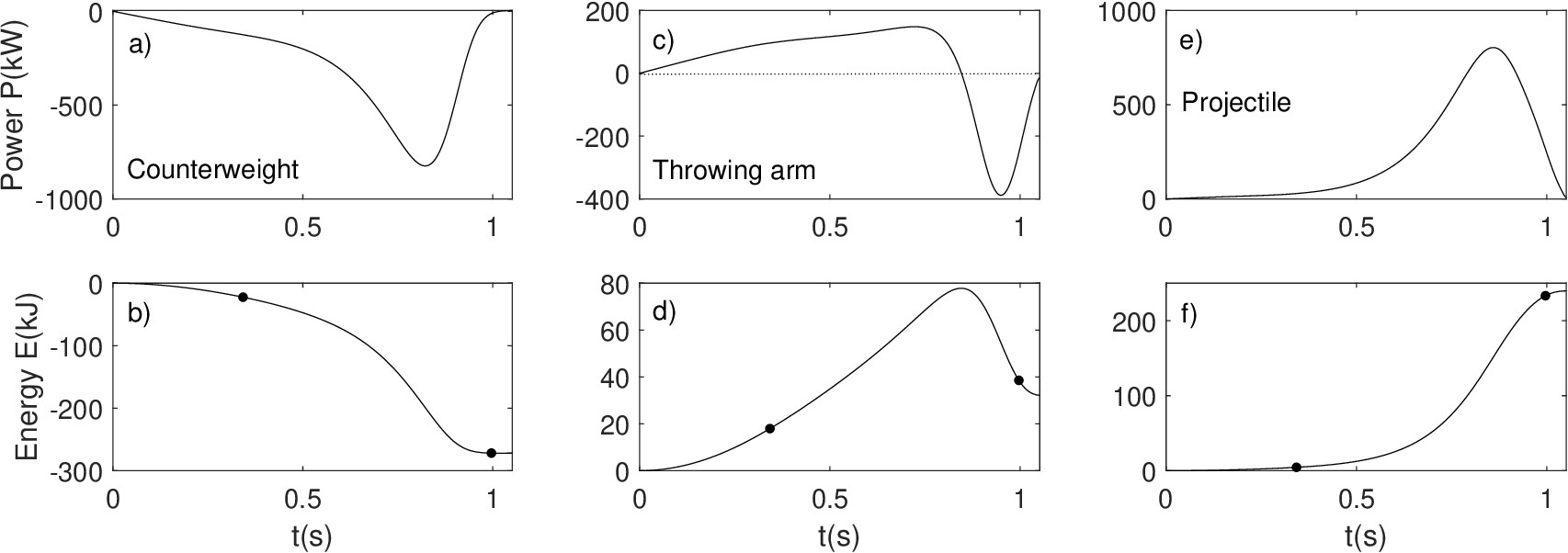}
	\caption{
		Upper panel:
		Rates of work by~$\Fm_H$ on counterweight,~$\Nm$ and gravity on throwing arm,
		and~$\Fm_S$ on projectile.
		Lower panel:
		Accumulated mechanical energies.}
	\label{fig:Power_Beam}
\end{figure}

The counterweight moves at the velocity~$\vm_M$, and the internal force~$\Fm_H$ 
that acts on it through the arm~$L_3$ is such that~$\Fm_H\cdot\vm_M<0$ 
at all times until release.
The power is then always negative~\eref{eq:Energy_M}, so the counterweight 
steadily looses mechanical energy as illustrated in~\fref{fig:Power_Beam}a 
and~\ref{fig:Power_Beam}b. 
The torque~$N$ and gravity at the center of mass determine the power on the arm 
according to~\eref{eq:dUdt_beam}.
As seen in~\fref{fig:Power_Beam}c and~\ref{fig:Power_Beam}d, the power is first positive 
and later negative, but the mechanical energy of the arm remains positive throughout
and reaches almost~80kJ or~30\% of the available mechanical energy~$\D U$ of~255kJ.
The sling tension~$\Fm_S$ and projectile velocity~$\vm_m$ determine the projectile 
power~\eref{eq:dEdt}.
\Fref{fig:Power_Beam}e and~\ref{fig:Power_Beam}f show that it is always positive,
so the mechanical energy of the projectile~\eref{eq:Energy_m} is steadily increasing
until it reaches~233kJ at release, which is more than~90\% of~$\D U$.

We have seen that the flow of mechanical energy from counterweight to projectile
is affected by the throwing arm.
During the first~85\% of a shot's duration, the arm stores mechanical energy flowing 
from the counterweight, but most of this energy is transferred to the projectile over
the remaining short period, so the rate becomes quite high and reaches almost~400kW.
When this is added to the continuing high power from the counterweight, the total 
increases to more than~800kW shortly before release. 
\section{Summery}
All forces that relate to a trebuchet are the sum of motion-dependent terms and
a constant vertical term from gravity.
Analytical expressions, which depend on angular coordinates and their derivatives 
to second order, are derived for the dynamic terms, which are cast into the form 
of mass times acceleration.
Discontinuities are seen at the moment a shot is initiated and as it progresses,
the dynamic terms become large.

The internal forces comprise, as examples, the force at the fulcrum 
and the sling tension.
In an illustrative design, the magnitude of the reaction force increases to~2.5 times the 
gravitational weight of the counterpoise, and in the same example, 
the tension rises to~52 times the gravity of the projectile.
Simple estimates of such forces can be misleading: 
Under the assumption of circular projectile motion at the release speed and with
radius equal to sling length, one finds a tension of~$81mg$ or an overshoot of~55\%
\footnote[1]{tension$/(mg)=(mv_r^2/L_4)/(mg)=66^2/(5.5\cdot9.82)=81$.}.

Internal forces are crucial for losses of mechanical energy and strengths of engine
components.
The most important losses are found at the bearings for the shaft that carries the
throwing arm and at the hinge for the counterweight.
Heat is generated here due to sliding friction at rates proportional to the
appropriate reaction forces and sliding speeds.
Short shafts with just sufficient diameters and strengths are essential for limiting 
the sliding speeds and therefore losses.
Friction also causes wear on the bearings, and the losses reduce range 
and projectile energy at target.

The required rigidity of the trestle that supports the engine depends on the magnitude 
of the reaction force at the bearings and the rapidly varying component in the 
horizontal direction. 
The same force equals the bending load on the throwing arm, 
which determines its diameter.

The flow of mechanical energy within the engine from counterweight to projectile
goes through the throwing arm.
It temporarily possesses~30\% of the energy, but the dominating kinetic part 
flows on to the projectile, which carries~91.3\% of the available mechanical 
energy at release in the ideal case without internal friction losses.

\newpage
\appendix
\section{Equations of motion in phase~I and initial accelerations}\label{app:EOM}
The equations for beam and counterweight motions in matrix form read
\begin{eqnarray}
	\left\{			
		\begin{array}{ccc}
			mL_1^2f(\ta)^2+ML_2^2+\I && -ML_2L_3\cos(\ta-\psi) \\
			-ML_2L_3\cos(\ta-\psi) 					&& ML_3^2	 
		\end{array}
	\right\}
	\left\{			
		\begin{array}{c}
			\ddot\ta   \\
			\ddot\psi   
		\end{array}
	\right\}= 															\label{eq:EOM}\\
	\left\{			
		\begin{array}{c}
			ML_2L_3\sin(\ta-\psi)\dot\psi^2-mL_1^2f(\ta)g(\ta)\dot\ta^2+(ML_2-m_bL_{CM})g\cos\ta \\
			-ML_2L_3\sin(\ta-\psi)\dot\ta^2 -ML_3g\cos\psi
		\end{array}
	\right\},
	\nonumber
\end{eqnarray}	
where~$f$ and~$g$ are defined in~\eref{eq:f_def} and~\eref{eq:g_def},
and~$\I$ is the moment of inertia of the beam with respect to the pivot,
and the projectile motion is determined by~\eref{eq:Const_I1}.

At rest initially with~$\ta=\ta_i$ and~$\psi=\psi_i=-\pi/2$, equation~\eref{eq:EOM} reduces to
\begin{eqnarray}
	\left\{			
		\begin{array}{ccc}
			mL_1^2\sin^2\ta_i+ML_2^2+\I && ML_2L_3\sin\ta_i \\
			L_2\sin\ta_i					&& L_3	 
		\end{array}
	\right\}
	\left\{			
		\begin{array}{c}
			\ddot\ta_i   \\
			\ddot\psi_i   
		\end{array}
	\right\}= 															\nonumber\\
	\left\{			
		\begin{array}{c}
			(ML_2-m_bL_{CM})g\cos\ta_i				\\
			0
		\end{array}
	\right\}.
	\label{eq:EOM_init}
\end{eqnarray}	
The determinant of the~$2\times2$ matrix in~\eref{eq:EOM_init} is
\begin{eqnarray}
		D=L_3\left(mL_1^2\sin^2\ta_i+ML_2^2\cos^2\ta_i+\I\right)>0,
		\nonumber
\end{eqnarray}
so the matrix can be inverted, and the initial accelerations are
\begin{eqnarray}	\left\{			
		\begin{array}{c}
			\ddot\ta_i   \\
			\ddot\psi_i   
		\end{array}
	\right\}
	&=\frac{1}{D}
	\left\{			
		\begin{array}{cc}
			L_3	 					& \dots \\
			-L_2\sin\ta_i	& \dots 
		\end{array}
	\right\} 							
	\left\{			
		\begin{array}{c}
			(ML_2-m_bL_{CM})g\cos\ta_i				\\
			0
		\end{array}
	\right\}												\nonumber\\
	&=\frac{(ML_2-m_bL_{CM})g\cos\ta_i}{D}
	\left\{		
		\begin{array}{c}
			L_3	  \\
			-L_2\sin\ta_i
		\end{array}
	\right\}.											\nonumber
\end{eqnarray}	
Differentiation of equation~\eref{eq:Const_I2} determines~$\ddot\phi_i$ by 
\begin{eqnarray}	
	\ddot\phi_i=\frac{L_1}{L_4}\cos\ta_i\ddot\ta_i
	. \nonumber
\end{eqnarray}	

\end{document}